\documentclass[apj,twocolumn, twocolappendix]{openjournal}
\usepackage{amsmath}
\usepackage{booktabs}
\usepackage{multirow}
\usepackage{color}
\usepackage{soul}
\usepackage{multirow}
\usepackage{array}
\usepackage{lastpage}

\usepackage[pdfencoding=auto,psdextra]{hyperref}
\hypersetup{
    colorlinks=true,
    linkcolor=blue,
    filecolor=magenta,      
    urlcolor=blue,
    citecolor=blue
}

\newcommand{\MBH}{$M_\mathrm{BH}$}
\newcommand{\MBHm}{M_\mathrm{BH}}
\newcommand{\Msun}{M$_\odot$}
\newcommand{\Msunm}{\mathrm{M}_\odot}
\newcommand{\MS}{$M_\star$}
\newcommand{\MSm}{M_\star}
\newcommand{\mSm}{m_\star}
\newcommand{\MBHMS}{$M_\mathrm{BH}$--$M_\star$}

\newcommand{\MBHDsm}{\rho_\mathrm{BH,\star}}
\newcommand{\MBHD}{$\rho_\mathrm{BH}$}
\newcommand{\MBHDs}{\rho_\mathrm{BH,\star}}
\newcommand{\MAG}{\mathcal{M}}

\usepackage[nameinlink,capitalise]{cleveref}
\crefname{section}{Sect.}{Sects.}
\Crefname{section}{Section}{Sections}
\crefname{figure}{Fig.}{Figs.}
\Crefname{figure}{Figure}{Figures}
\crefname{equation}{Eq.}{Eqs.}
\Crefname{equation}{Equation}{Equations}
\crefname{table}{Table}{Tables}
\crefname{appendix}{Appendix}{Appendices}

\renewcommand{\arraystretch}{1.2}  
\usepackage{orcidlink}

\begin{document}

\title{The Soltan argument at redshift 6:\\ UV-luminous quasars contribute less than 10\% to early black hole mass growth}

\shorttitle{The Soltan argument at $z=6$}
\shortauthors{K.\ Jahnke}

\author{Knud Jahnke\,\orcidlink{0000-0003-3804-2137}$^{1}$}

\affiliation{$^1$Max Planck Institute for Astronomy, K\"onigstuhl 17, 69117 Heidelberg, Germany}
\email{jahnke@mpia.de}

\begin{abstract}
We combine stellar mass functions and the recent first JWST-based galaxy--black hole scaling relations at $z=6$ to for the first time compute the supermassive black hole (SMBH) mass volume density at this epoch, and compare this to the integrated SMBH mass growth from the population of UV-luminous quasars at $z>6$.
We show that even under very conservative assumptions almost all growth of supermassive black hole mass at $z>6$ does not take place in these UV-luminous quasars, but must occur in systems obscured through dust and/or with lower radiative efficiency than standard thin accretion disks.
The `Sołtan argument' is not fulfilled  by the known population of bright quasars at $z>6$: the integrated SMBH mass growth inferred from these largely unobscured active galactic nuclei (AGN) in the early Universe is by a factor $\ge$\,10 smaller than the total black hole mass volume density at $z=6$. 
This is valid under a large range of assumption about luminosity and mass functions as well as accretion modes, and is likely still at least a factor $>$\,2 smaller when accounting for the known obscuration fractions at this epoch. 
The resulting consequences are: $>$\,90\%, possibly substantially more, of SMBH-buildup in the early Universe does not take place in luminous unobscured quasar phases, but has to occur in obscured systems, with dust absorbing most of the emitted UV--visible AGN emission, potentially with accretion modes with super-Eddington specific accretion rates. This is consistent with short lifetime arguments for luminous quasar phases from quasar proximity zone studies and clustering. This would remove the empirical need for slow SMBH growth and hence exotic `high-mass seed' black holes at early cosmic time. It also predicts a large population of luminous but very obscured lower-mass quasars at $z>6$, possibly the JWST `Little Red Dots'. 
\end{abstract}


\maketitle

\section{Introduction} \label{sec:intro}

Quasars -- accreting supermassive black holes (SMBHs) -- have been observed throughout cosmic time, all the way back to (currently) a redshift of $z\sim7.5$, an epoch when the Universe was merely 700\,Myr old. Ongoing projects for example with JWST datasets \citep[e.g.\ COSMOS-Web][]{casey2023,lambrides2024,andika2024} as well the systematic survey by {\it Euclid} \citep{euclid2024,barnett2019,selwood2024} have the aim to extend this range by a few 100\,Myr closer to the Big Bang, aiming ultimately for $z>9$. These are all searches for broad-line quasars,  largely unobscured by dust and selected at rest-frame UV--visible wavelengths, which provide a comparably easy way of identification as well as estimating their SMBH masses from single epoch spectra using their broad emission lines.

What this, currently known, quasar population in the early Universe has in common is that their SMBHs have substantial masses. All of them have $\MBHm > 10^7$\,\Msun, a fraction even beyond $10^9$\,\Msun \citep[e.g.][]{mortlock2011,banados2018, banados2023,yang2020,takahashi2024}. This population is the origin of the long-identified timescale and `black hole seed problem': with the observed Eddington-limited accretion \citep[e.g.][]{takahashi2024} the time needed to grow these SMBH from low-mass -- $\le$\,100\,\Msun\ -- seed black holes is longer than the age of the Universe at that epoch \citep{banados2018,yang2020}.

The main proposed solution to this conundrum is the postulation of `high-mass SMBH seeds' \citep[e.g.][]{volonteri2010,inayoshi2020,volonteri2021}. Currently, proposed high-mass seed solutions like runaway stellar mergers in very dense clusters or the direct collapse of mass into SMBHs seem to all have drawbacks, mostly by requiring very specific and hypothetical physical conditions or mechanisms that have not been encountered at later cosmic epochs -- and hence do not have an empirical basis (yet).

For this reason we want to take a step back and ask the question: {\it How relevant is the currently known population of luminous UV--visible quasars actually for our learning about early SMBH growth?} This is deliberately phrased provocatively and could be seen as contentious. However, we do know from studies ranging from X-ray to radio studies \citep[e.g.][]{vito2018,gilli2022,caccianiga2024,satyavolu2023,banados2024} that a part of the SMBH growth in the very early Universe does not take place in these known, unobscured quasars with approximately Eddington accretion. It occurs rather in more obscured galaxy centers for which we currently can not determine SMBH masses and accretion rates properly -- or at all. In very strongly obscured galaxies there is the challenge to attribute observed emission to SMBH growth, and with any type of obscuration that prevents SMBH mass measurements, the {\it specific} SMBH growth rate and therefore timescale of growth evades empirical investigation if only X-ray and infrared properties of the SMBH are available. If a substantial fraction of early SMBH growth would occur in different systems from those we currently know and study at high-$z$, how far can diagnostic of these known quasars take us? And therefore: Are we in the end really asking the right questions?

In the following we hence make a test of whether the known SMBH growth in broad-line quasars observed at $z>6$ can actually explain the build-up of all the SMBH mass in galaxies at $z=6$, or if there is a major discrepancy that requires a lot of this growth to currently not be accounted for by the known luminous quasar population. This is in principle the `Sołtan argument' \citep{soltan1982}, but now at $z=6$ instead of the local Universe. We focus on the quasars selected and characterised at rest-frame UV--visible wavelengths, since currently we draw most diagnostics of early SMBH growth from these. We already know that there is SMBH growth in obscured quasars, as we know from the lower-redshift Universe. But it is not immediately clear whether these typically X-ray selected quasars, often with lower SMBH masses and lower accretion rates. Hence their importance at $z>6$ is not clear, since data is scarce, and -- quite importantly -- it is not clear if these known type-2 quasars are the only population we might be missing in the first billion years.

The Sołtan argument requires two ingredients, a total SMBH mass census at $z\sim6$ one side and the integral of SMBH accretion turned into \MBH\ on the other\footnote{We note that this is a question of mass volume density buildup and is not about the formation of the SMBH mass {\it function}.}. While the latter has in principle already been available for a while, the former has not. The reason was that while in the later Universe we could easily either directly measure SMBH in galaxies from dynamics or orbit modelling \citep[e.g.][]{haring2004}, in the early Universe we either have access to stellar mass but not SMBH mass in nuclear-inactive galaxies, or SMBH mass but not stellar mass for UV-luminous quasars. So the relative mass scales -- this is the characteristic scaling relation converting \MS\ to \MBH\ -- was observationally only available at lower redshifts \citep{jahnke2009,cisternas2011,sun2015,tanaka2024}.

This recently changed with JWST. With these \MBHMS\ scaling relations now available we can for the first time calculate the volume density of SMBH mass at $z\sim6$, as scaled from the well established stellar mass functions of galaxies. A comparison with the SMBH mass accreted in UV-luminous quasars will then tell us how important these quasars are in the overall SMBH mass growth at early times, and whether questions that specifically arise from them -- for example the seed problem -- are indeed the right questions to ask. And, as a byproduct, whether we need to focus more on other wavelengths, selection methods, and diagnostics in order to understand early SMBH growth.

With that said, the approach that we are taking in the following is rather straightforward and is based on empirical information about SMBH mass and SMBH mass growth: We will compare the integral of visible SMBH growth -- as represented by the known population of UV-luminous, largely unobscured broad-line quasars -- between the big bang and $z=6$, with the independently inferred overall SMBH mass density at $z=6$ as inferred from stellar mass functions and the new \MBHMS\ relations.

We start with inferring the integrated SMBH growth to $z=6$, then calculate the total SMBH volume density at that redshift, and then compare the two. This includes and critically hinges on a discussion of our choices of assumptions and a sensitivity analysis for other conservative and non-conservative parameter choices.

While most of the calculations are related to comoving volumes, when cosmic time is needed we are using the Planck 2018 cosmology \citep{planck2020}, with $H_0=67.7$\,km/s/Mpc, $\Omega_m=0.310$, and a flat geometry. However our results only change on the 1\%-level if other cosmologies are used.
\bigskip


\section{The integrated SMBH mass growth to \texorpdfstring{$z=6$}{z=6}}

The first ingredient to the Sołtan argument is the predicted SMBH mass (or rather mass volume density) as calculated from visible SMBH growth. 

As stated above, we focus on the general population of UV-luminous quasars at $z>6$ and the quasar luminosity functions (QLFs) that have been derived for them. For a time-integral of SMBH growth since the big bang we add the evolution of the QLF to earlier times. Since others have done this before using different input populations, we will juxtapose the resulting values for \MBHD\ to \citet{shen2020}, who have folded in both a full bolometric view as well as assumptions about the fraction of obscured and absorbed SMBH growth.


\subsection{Quasar luminosity function at \texorpdfstring{$z=6$}{z=6}}\label{sec:qlf}

As the input QLF, $\Phi(\MAG) d\MAG$, denoting the comoving number density of of quasars in the magnitude range $\MAG$ to $\MAG+d\MAG$, with the functional form

\begin{equation}
    \Phi(\MAG) = \frac{\Phi^*}{10^{0.4(\alpha+1)(\MAG-\MAG^*)}+10^{0.4(\beta+1)(\MAG-\MAG^*)}}\,, 
\end{equation}
we are using two different estimates: One is the $z=6$ QLF at 1450\,\AA\ QLF by \citet{schindler2023}, from here on `S23', based on SHELLQs \citep{Matsuoka2018} and Pan-STARRS1 \citep{banados2016} with
\begin{equation}
    \begin{split}
    \log(\Phi^*/(\mathrm{mag}^{-1}\,\mathrm{Mpc}^{-3}))&=-8.75^{+0.47}_{-0.41}, \\
    \MAG^*&=-26.38^{+0.79}_{-0.60}\,\mathrm{mag}_{AB},\\
\alpha&=-1.70^{+0.29}_{-0.19}, \quad \mathrm{and}\\ 
\beta&=-3.84^{+0.63}_{-1.21}.
    \end{split}
\end{equation}
The second is by \citet{Matsuoka2018} directly, from here on `M18', with
\begin{equation}
    \begin{split}
\log(\Phi^*/(\mathrm{mag}^{-1}\,\mathrm{Mpc}^{-3}))&=-7.96^{+0.32}_{-0.042},\\ 
\MAG^*&=-24.9^{+0.75}_{-0.9}\,\mathrm{mag}_{AB},\\
\alpha&=-1.23^{+0.44}_{-0.34}, \quad \mathrm{and}\\ 
\beta&=-2.73^{+0.23}_{-0.31}.
    \end{split}
\end{equation}

We convert these QLFs from magnitude to linear units, then multiply with luminosity, apply a bolometric correction, and finally integrate over all luminosities from $L_\mathrm{low}$ to $L_\mathrm{high}$ to get the luminosity (volume) density $\rho_L$. 

For the bolometric correction (BC), converting from the luminosity at 1450\,\AA\ to bolometric luminosity, we test the two different forms by \citet{runnoe2012,runnoe2012b}, one with constant $\mathrm{BC}_\mathrm{1450\,\textnormal{\AA}} = 4.2$, the other in the form $\log (L_\mathrm{bol} = 4.745 + 0.91\log(1450\,\textnormal{\AA}\, L_\textnormal{1450\,\AA})$. For the M18 QLF the resulting $\rho_L$ is almost independent from the choice of BC at all $L$ cutoffs. For the S23 QLF with its steeper faint-end slope there is a mild dependency.

\begin{figure}[htb]
   \includegraphics[width=1.0\columnwidth]{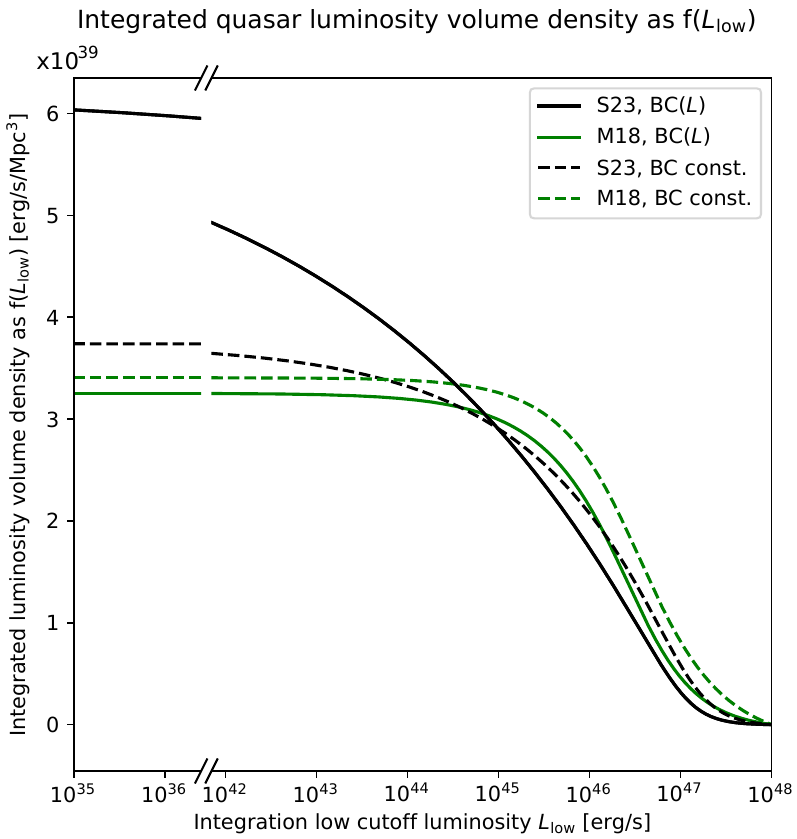}
    \caption{Bolometric quasar luminosity volume density for largely unobscured UV-luminous quasars at $z=6$, as function of lower cutoff luminosity for integrating the QLF. Black lines show the values for the S23 QLF, green lines for the M18 QLF. Dashed lines use a constant bolometric correction, solid lines a luminosity-dependent BC. For $L_\mathrm{low}\rightarrow 0$ the S23 line converges to $\sim$6\,$\times$\,10$^{39}$\,erg/s/Mpc$^3$.}
    \label{fig:lumdensity}
\end{figure}

The integral for $\rho_L$ is insensitive to the choice of $L_\mathrm{high}$, due to the steep power-law at high luminosities. Beyond $\log (L/(\mathrm{erg}\,/\mathrm{s})) = 47$ the contributions to $\rho_L$ become almost negligible. The dependency on the faint end cutoff choice $L_\mathrm{low}$ is shown in \cref{fig:lumdensity}. Since both M18 and S23 directly cover the QLF with data about two orders of magnitude below $\MAG^*$, that is to $L_\mathrm{bol}\sim 10^{44}$\,erg/s, we assume that their model holds at least another dex fainter without too much error.

With this similarity of integrated $\rho_L$ we choose to adopt the S23 QLF with the $L$-dependent BC, as this provides the highest value at low cutoffs -- conservative for our discussion goal -- but generally numbers are not that different from the M18 QLF.
The resulting  $\rho_L$ are: for $L_\mathrm{low}=10^{44}$\,erg/s the integrated quasar $\rho_L(>10^{44}\,\mathrm{erg/s})=3.8\times10^{39}$\,erg/s/Mpc$^3$, for lower cutoffs this becomes $\rho_L(>10^{43}\,\mathrm{erg/s})=4.4\times10^{39}$\,erg/s/Mpc$^3$ and $\rho_L(>10^{42}\,\mathrm{erg/s})=4.9\times10^{39}$\,erg/s/Mpc$^3$, respectively.


\subsection{Radiated energy since \texorpdfstring{$z=\infty$}{z=infty}}
\label{sec:evobol}

In order to get to a total energy radiated by broad-line quasars in the early Universe we now need to integrate over redshift or rather time, to take care of the evolution of the QLF. Without QLF evolution the integration would revert to multiplying with the cosmic age, that is 929\,Myr for $z=6$ in a Planck 2018 cosmology.

\begin{figure}[htb]
   \includegraphics[width=1.0\columnwidth]{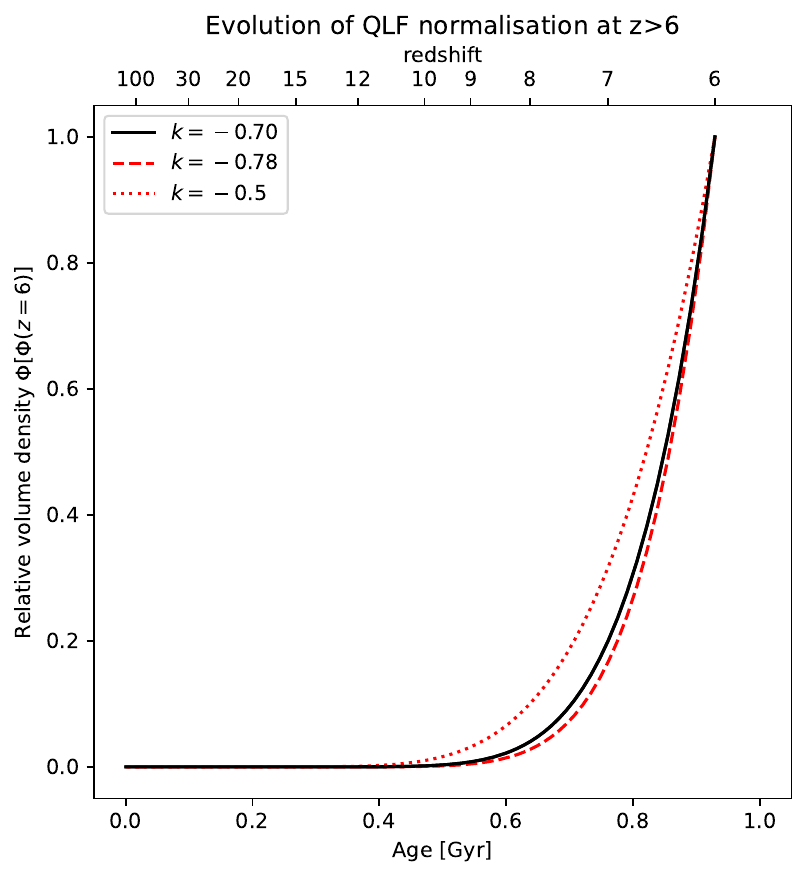}
    \caption{Fast evolution of the quasar luminosity function before $z=6$. The curves are normalised to 1 at $z=6$ and show the evolution for three different parameters $k$ as $(1+z)^k$, with values $k=-0.70$ (solid line), $k=-0.78$ (dashed line), and an extreme $k=-0.5$ (dotted) line as a general comparison. The area under the curve is the effective time multiplier to convert the luminosity volume density at $z=6$ into a cumulated emitted energy by quasars from $z=\infty$ to $z=6$.}
    \label{fig:lfevo}
\end{figure}

But the QLF is indeed strongly evolving. With the typical parametrisation $LF(z)= LF(z_0)\times10^{k(z-z_0)}$ there is a consensus that evolution is fast in the early Universe, \citet{schindler2023} use $k=-0.70$, \citet{wang2019} find $k=-0.78$ and much slower evolutions are not being proposed in the literature. In \cref{fig:lfevo} we show this evolution, the value $k=-0.70$ as a solid line, $k=-0.78$ dashed, and for illustration purposes an extreme $k=-0.5$ -- we note this has not been suggested in the literature for these redshifts -- simply to gauge the impact.

If we integrate over time including this evolution, $t_\mathrm{evo}$, the QLF at $z=6$ has to be multiplied with a time $t_\mathrm{evo}(k=-0.70) = 103$\,Myr. For $k=-0.78$ this decreases by 9\% to $t_\mathrm{evo}(k=-0.78) = 94$\,Myr, and for the extremely slow evolution of $k=-0.5$ this would increase by $\sim$\,30\% to $t_\mathrm{evo}(k=-0.5) = 135$\,Myr.

This results in a conservative total energy emitted from UV-luminous broad-line quasars at $z\ge6$
\begin{equation}
 E_{\mathrm{quasars},z\ge6} = \rho_L(>10^{42}\,\mathrm{erg/s}) \times t_\mathrm{evo}(k=-0.70)
\end{equation}
and a value $E_{\mathrm{quasars},z\ge6} = 1.6\times10^{55}$\,erg/Mpc$^3$.


\subsection{Conversion to SMBH mass at \texorpdfstring{$z=6$}{z=6}}\label{sec:quasarmass}

The final step is the conversion from radiated energy to mass. This is possible under the combination of mass--energy equivalence \citep{einstein1905} and assumption of a specific radiative efficiency $\varepsilon$ of accretion disks, that is a specific fraction of infalling mass being converted into radiative energy, while only the remainder falls into the black hole.

For a standard thin disk model the efficiency takes a value of $\varepsilon\sim0.1$ for sub-Eddington accretion rates \citep{shakura1973,abramowicz2013}, with a decrease for very high accretion rate. The quasars we are concerned with are all radiating at substantial fractions of the Eddington accretion rate where a value of $\varepsilon=0.1$ should both be appropriate and provides a conservatively high value for the resulting mass density. 

With this the conversion from cumulative bolometric radiated energy volume density and accreted SMBH mass density $\rho_\mathrm{BH,acc}$ becomes
\begin{equation}
    \rho_\mathrm{BH,acc} = \frac{E_{\mathrm{quasars},z\ge6}}{c^2}\frac{(1-\varepsilon)}{\varepsilon}
\end{equation}
with $c$ being the speed of light. This converts the previously calculated $E_{\mathrm{quasars},z\ge6} = 1.6\times10^{55}$\,erg/Mpc$^3$ into 
\begin{equation}
    \rho_\mathrm{BH,acc} = 79\,\Msunm/\mathrm{Mpc}^3 
\end{equation}
by $z=6$.


\section{The total SMBH mass density at \texorpdfstring{$z=6$}{z=6}}
\label{sec:MBHdensity}

\subsection{\texorpdfstring{\MBH}{MBH}--\texorpdfstring{\MS}{MS} scaling relations at \texorpdfstring{$z=6$}{z=6}}\label{sec:scalrel}

While at $z\sim6$ we have known for 25 years an increasing number of quasars and subsequently their SMBH masses \citep[e.g.][]{fan2001}, we could never estimate the total SMBH mass in the whole of the galaxy population at this epoch. Now we can.

This is made possible by the first estimates of the SMBH mass--stellar mass scaling relations at $z\sim6$ using JWST. Already the first two quasar host galaxies whose masses were measured with JWST \citep{ding2022} enabled an estimate of the underlying \MBHMS\ scaling relations, when factoring in all selection and scatter effects implicit in the observation of these quasars -- obviously still with substantial uncertainties. These shrank with a number of projects subsequently providing more datapoints and the currently most robust and detailed estimate of the \MBHMS\ relation relation is a meta-analysis by \citet{li2024a}, including also the crucial faint end of the quasar population, using data by \citet{ding2022},  \citet{harikane2023}, \citet{maiolino2024}, \citet{stone2024}, and \citet{yue2024}. They parametrise the relation as
\begin{equation}\label{eq:scalrel}
\MBHm/\Msunm = a \times \log (\MSm/\Msunm) + b
\end{equation}    
and estimated $a=1.07^{+0.57}_{-0.57}$ and $b=-3.75^{+5.08}_{-5.35}$. Here \MS\ is the galaxy total stellar mass, \MBH\ the SMBH mass. The underlying sample of 32 quasars were all observed with JWST and their redshifts stretch from $4<z<7$ with most of them lying around $z=6$. These scaling relations lie even slightly lower in \MBH\ than the initial result by \citet{ding2022}, but both results are consistent with the scaling relation at $z=0$ within a few 0.1\,dex. This result is new and was not necessarily expected, as it demonstrates that the relative efficiency of SMBH and stellar mass growth is similar in the early and later Universe -- but the impacts of this are subject to a discussion elsewhere. We just want to note here that the \citet{li2024a} scaling relation that we will build on in the following is {\it conservative} with respect to predicted \MBH, it predicts low BH masses. Any higher normalisation -- as some studies would prefer \citep[e.g.][]{pacucci2023,stone2024,maiolino2024} -- would mean a higher \MBH\ by the same amount, and hence an inferred higher BH mass density $\MBHDsm$ at $z=6$.


\subsection{Stellar mass function and derived SMBH mass function at \texorpdfstring{$z=6$}{z=6}}\label{sec:massfunction}

With this relation established, it can be used to convert the stellar mass function (MF) at $z=6$, which has been estimated for a few years already \citep{stefanon2021}, into an overall SMBH mass function that includes SMBHs in all active and inactive galaxies. The underlying assumption is that by $z=6$ the SMBH halo occupation fraction should be very high, meaning most galaxies should already have a supermassive black hole in their center, since hierarchical structure formation \citep{jahnke2011} has been at work for a several dynamical timescales at that epoch: massive galaxies are predicted to having had one or several major mergers by that time \citep{oleary2021}, contributing $\ge$\,10\% of their stellar content \citep{rodriguez2016}.

We want to base our calculations on two stellar MFs at $z=6$ by \citet{stefanon2021}, `ST21' in the following, using the deepest Spitzer IRAC 3.6\,\micron\ and 4.5\,\micron\ photometry data available for $\sim$800 Lyman-break galaxies at $6<z<10$, and by \citet{weibel2024}, `WE24' in the following, based on JWST NIRCam data of 30,000 galaxies at $4<z<9$. Both provide a Schechter-function fit to their data
\begin{equation}\label{eq:stellarmassfct}
\phi_\star(M) = \ln(10) \phi^*_\star 10^{(\mSm-\mSm^*)(1+\gamma)}\exp(-10^{(\mSm-\mSm^*)}).
\end{equation}
With $\mSm = \log (\MSm/\Msunm)$ and $\mSm^* = \log (\MSm^*/\Msunm)$, S21 find these best-fit parameters for low-mass-end slope $\gamma$, characteristic stellar mass $\MSm^*$, and characteristic volume density $\phi^*_\star$
\begin{equation}
    \begin{split}
        \gamma&=-1.88^{+0.06}_{-0.03},\\
        \log(\MSm^*/\Msunm)&=10.24^{+0.08}_{-0.11},\quad \mathrm{and}\\
        \log(\phi^*_\star/\mathrm{dex}^{-1}/\mathrm{Mpc}^{-3}) &= -4.09^{+0.17}_{-0.12},
    \end{split}
\end{equation}
while WE24 find
\begin{equation}
    \begin{split}
        \gamma&=-1.95^{+0.08}_{-0.06},\\
        \log(\MSm^*/\Msunm)&=10.01^{+0.28}_{-0.36},\quad \mathrm{and}\\
        \log(\phi^*_\star/\mathrm{dex}^{-1}/\mathrm{Mpc}^{-3}) &= -4.26^{+0.36}_{-0.36}.
    \end{split}
\end{equation}

Please note that we will not use formal uncertainties in the following calculations, but will focus in Sect.~\ref{sec:sensitivity} on the dominating integration boundaries of mass and luminosity functions. 

\begin{figure}[htb]
   \includegraphics[width=1.0\columnwidth]{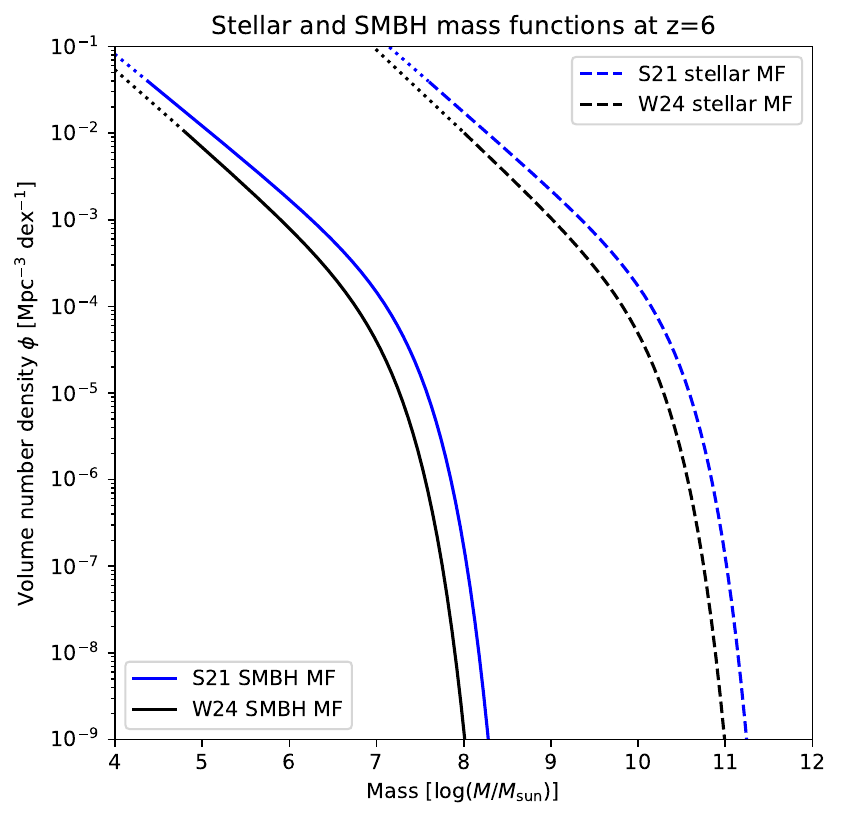}
    \caption{Stellar mass functions (dashed lines) and inferred SMBH mass functions (solid lines). Blue lines refer to the ST21 MF, black lines to the WE24 MF. At the high mass end we note that in both MFs \MS\ is only constrained by data out to $\log(\MSm/M_\odot)\sim10.6$. While the massive end does not contribute substantially to the total mass, the notable existence of $\log(\MBHm)\sim9$ quasars at $z=6$ is not at odds with this. At the low mass end the dotted portions of the mass functions are extrapolations beyond mass ranges with complete data in the respective studies.}
    \label{fig:massfncts}
\end{figure}

These stellar MFs are shown as dashed lines on the right of \cref{fig:massfncts}. Substituting \cref{eq:scalrel} into this stellar MF of \cref{eq:stellarmassfct} derives the corresponding SMBH mass function $\phi_\mathrm{BH}(M)$, shown as the solid lines on the left of \cref{fig:massfncts}. Here the dashed and solid lines mark ranges where these studies have mass-complete data, while the dotted sections at low masses are extrapolations of the parametrised functions beyond this range.

\begin{figure}[htb]
   \includegraphics[width=1.0\columnwidth]{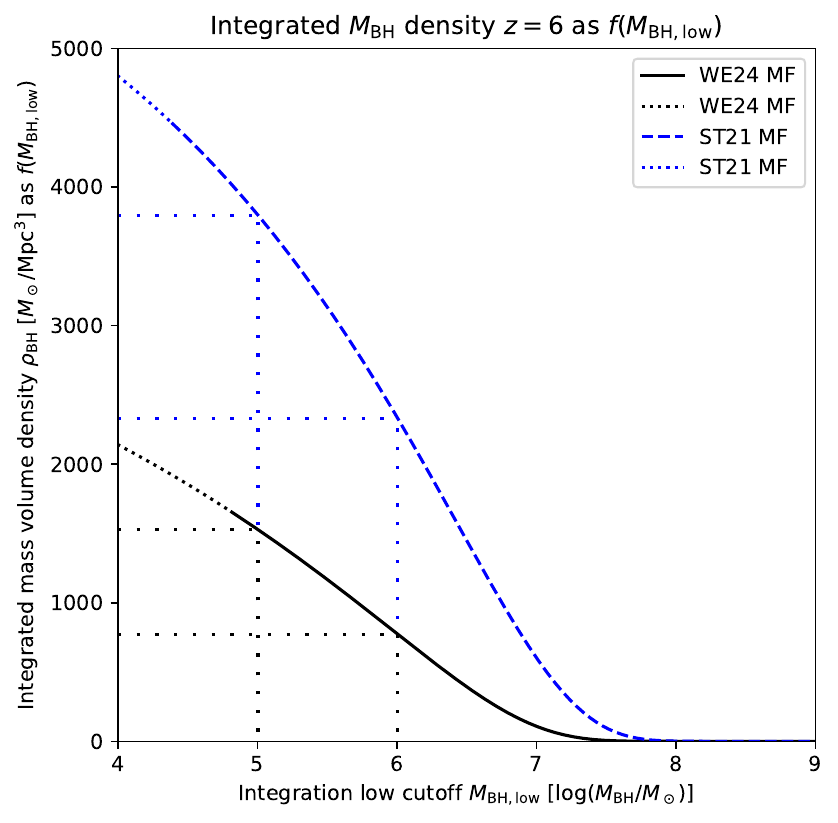}
    \caption{Integrated SMBH mass volume density as derived from galaxy stellar MFs and the \MBHMS\ scaling relation at $z=6$, as function of the lower cutoff mass for integrating the MF. The blue line is for the ST21 mass function, the black line for the WE24 mass function, dashed/solid sections are covered by mass-complete data in S21/W24.}
    \label{fig:massdensity}
\end{figure}

The total SMBH mass density derived from this mass function will depend on the integration boundaries. For our comparison in principle only a certain \MBH\ mass range will be observable as quasars with current data. However, given information about luminosity function shapes from lower redshifts, and the fact that the observed quasar luminosity is a combination of \MBH\ and specific accretion rate, the lower limits integration limits are not a priori clear. As a basis for a later sensitivity analysis and discussion in Sects.~\ref{sec:sensitivity} and \ref{sec:discussion} the resulting \MBH\ volume density 
\begin{equation}
\MBHDsm = \int_{\MBHm,\mathrm{low}}^{10^9\Msunm}  \phi_\mathrm{BH}\MBHm d(\MBHm)
\end{equation}
as a function of the lower integration mass limit $\MBHm,\mathrm{low}$ is shown in \cref{fig:massdensity} and tabulated in \cref{tab:mbh_ms}. All lower bounds that are being considered in \cref{tab:mbh_ms} are covered by mass-complete data in S21 and W24. The upper bound at the same time is not that important as the exponential drop of the Schechter function basically provides zero contributions from $\log(\MBHm/\Msunm)>8$. We see substantial differences by factors of 2--3, due to the lower normalisation of the WE24 MF. 

\begin{table}
    \centering
\begin{tabular}{p{0.2\columnwidth}>{\centering}p{0.2\columnwidth}>{\centering\arraybackslash}p{0.2\columnwidth}}
\hline
\rule{0pt}{3ex}  
\multirow{2}{*}{$\MBHm,\mathrm{low}$}  \rule{0pt}{3ex}  
&\multicolumn{2}{c}{$\MBHDs$\, [\Msun\,Mpc$^{-3}$]}\\
\cline{2-3}\rule{0pt}{3ex}  
&  ST21 & WE24\\
       \midrule
10$^6$\,\Msun &2330  & 770\\
10$^5$\,\Msun &3800  & 1530\\
    \end{tabular}
    \caption{Black hole mass volume density $\MBHDs$ at $z=6$ based on two different stellar mass functions and for two different low-mass integration cutoffs.}
    \label{tab:mbh_ms}
\end{table}


\section{Sensitivity analysis}
\label{sec:sensitivity}

Before we are discussing the (in-)consistency of these two SMBH mass density estimates, we need to dive into the uncertainties and dependencies associated with these calculations. We decided to do this assumption by assumption and not combine formal confidence intervals. This is because some of the assumptions have different types of uncertainties that require a discussion, mainly the choice of sources for the relations used, and the faint- and -- crucially -- low-mass-end cutoffs of the luminosity and mass functions.

\subsection{SMBH mass density from integrated quasar luminosity}

The choice of UV quasar luminosity function is rather unproblematic; for a comprehensive view both fainter and brighter quasars are necessary. The faint end almost automatically has to include the quasars discovered as part of the SHELLQs survey \citep{Matsuoka2018} using the Subaru telescope. Their original M18 QLF has a slightly flatter faint-end slope than the S23 QLF using PS1 and SHELLQs quasars, as reported above, but also a different normalisation. The impact on the integrated quasar luminosity volume density (\cref{fig:lumdensity}) is small, leading to $\sim$\,15\% higher values for $\rho_L(>10^{44}\,\mathrm{erg/s})$ and $\sim$\,50\% higher for $\rho_L(>10^{42}\,\mathrm{erg/s})$. 

The integral over the quasar emitted luminosity volume density is only weakly depending on luminosity cutoff for M18 and is still only weak for S23, due to the faint-end slope of $\alpha=-1.70$ for S23, as shown in \cref{fig:lumdensity}. This means our results are rather insensitive to a specific cutoff choice.
We conclude that the specific choice of $z=6$ QLF will only have a small impact on our interpretation and that choosing S23 and a cutoff at $10^{42}\,\mathrm{erg/s}$ will be conservative for our purpose. It is noteworthy that the faint-end slope is quite robust -- it is in line with other studies at similar redshifts -- and shows consistency with a slight steepening trend from lower redshifts, as shown in S23.  
We already also showed the good knowledge and limited impact of the specifics of QLF evolution (\cref{fig:lfevo}), with an impact on the final SMBH mass density of the range of evolution parameters that has been put forward on the order of not more than 10\%.

Two more assumptions entered our calculation: bolometric correction and radiative efficiency. Above we assumed two variants of the BC, one constant and one luminosity-dependent from the modelling by \citet{runnoe2012}. The latter has an increasing BC towards lower $L$ -- however even for $L_\mathrm{1450\,\textnormal{\AA}}=10^{43}$\,erg/s the calculated BC only rises to 7.5 from 3.3 at $L_\mathrm{1450\,\textnormal{\AA}}=10^{47}$\,erg/s. This is fully consistent with later calculations by \citet{shen2020} who arrive at very similar values. As a result, indicated by \cref{fig:lumdensity} the differences between these choices of BC are mild for both QLFs.

The radiative efficiency is a central ingredient, as it converts the luminosity proxy into SMBH growth. We are using the canonical $\varepsilon=0.1$ for non- or slowly spinning SMBHs, which is in line with other studies \citep{shen2020}. Only if the SMBH spin is starting to get closer to a Kerr solution we would need to assume a larger $\varepsilon$, which in turn would lead to a smaller value of $\rho_\mathrm{BH,acc}$. Lower values of $\varepsilon$ are only expected for different accretion disk configurations which we discuss below. Our choice of $\varepsilon$ is therefore conservative for our specific goal.

\subsection{SMBH mass density from stellar mass function}

The main assumptions in the computation of $\MBHDs$ at $z=6$ as inferred from the stellar mass function are, obviously, the stellar mass function, the \MBHMS-scaling relation, and the cutoff $M_\mathrm{BH,low}$ in integrating the SMBH mass function. On the stellar MF, the results of ST21 are based on ultra-deep Spitzer infrared data. Now the first deep JWST studies are available, although they are still constrained to rather small areas on the sky, not yet larger than the ST21 Spitzer area, but already with many more galaxies to work with. WE24 used JWST/NIRCam data on $\sim$\,500\,arcmin$^2$ to estimate the stellar MF at $4<z<9$. Their best-fit $z=6$ Schechter model has a low-madd-end slope $\gamma$ broadly consistent with ST21, but a $\sim$\,0.2\,dex lower $M^*_\star$ and 0.2\,dex lower $\phi^*_\star$. Correspondingly, their fit produced integrated stellar masses -- and therefore predicted SMBH masses -- that are smaller than those of ST21 by 67\% and 60\% for $10^5\,\Msunm$ and $10^6\,\Msunm$ low mass integration cutoffs (see \cref{tab:mbh_ms}). We note that for the small JWST area off 500\,deg$^2$ the number of massive galaxies beyond $\MSm\sim10^{10.0}$\,\Msun\ at $z=6$ is only a handful, so uncertainties of the stellar MF around $M^*_\star$ are substantial. However, a degeneracy in their model fit between $M^*_\star$ and $\phi^*_\star$ does not result in dramatic variations of integrated stellar mass density within their confidence region.

The second ingredient is the \MBHMS\ relation. The first ever scaling relation of this kind was only made possible by the first two quasar host galaxies at $z\sim6$ accessed with JWST \citep{ding2022}. 
The scaling relation by \citet{li2024a} uses more and also lower-$M$ quasar host galaxies -- also including the two objects by \citet{ding2022} -- and also accounts for the selection effects of all samples used. In the future the samples of high-$z$ quasar host galaxies will surely grow, but until then the \citet{li2024a} scaling relation will be the state of the art. There is possibly room for a higher normalisation of this relation, just judged from the fact that this currently lies even {\em below} the $z=0$ relation. However, with its -- fully data based -- low normalisation $b$ it is also conservative with respect to the inferred \MBH. An increase of $b$ by, say, 1\,dex would directly increase the predicted \MBH\ by a factor 10. So using the \citet{li2024a} scaling relations in combination with the stellar mass function by \citet{weibel2024} will result in the most conservatively low estimate of $\rho_\mathrm{BH,\star}$ currently available.

This only leaves two points, both related to the low-mass cutoff for defining $\rho_\mathrm{BH,\star}$. The first of these is the confidence of how low the stellar and hence derived BH mass functions are likely valid. \citet{weibel2024} are using data points as low as $\MSm = 4\times10^6$\,\Msun, corresponding to $\MBHm<10^4$\,\Msun, while stating mass-completeness down to $\MSm = 10^8$\,\Msun, corresponding to just below $\MBHm=10^5$\,\Msun. Given that intermediate mass black holes are only starting to be confirmed in the local Universe \citep{haeberle2024}, and that we do not know at which lower mass limit the \MBHMS-relations might break down, we clearly do not want to extrapolate too low, and want to select a conservative low cutoff $\MBHm=10^6$\,\Msun, which might be a good lower confidence bound for their validity in the high-$z$ Universe.

The second point is how to match the mass growth in quasars that we assess for $\rho_\mathrm{BH,acc}$ and the masses that we include in $\rho_\mathrm{BH,\star}$. Which we will discuss in the following.


\section{Discussion and impact}
\label{sec:discussion}

\subsection{Low mass cutoff and overlap of samples}
\label{sec:lowmasscutoff}

Almost exclusively, the $\sim$300 known UV-selected quasars at $z>5.8$ are the only population that currently informs us about SMBH masses and specific accretion rates in this epoch, i.e.\ is also at the heart of the `seed problem' of SMBH formation.
The underlying challenge in constructing or rather comparing these two black hole mass densities $\rho_\mathrm{BH,acc}$ and $\rho_\mathrm{BH,\star}$ is that these 300 quasars are only the tip of the iceberg of black holes, drawn from the whole sky, while the available stellar mass functions, that we use as a basis for $\rho_\mathrm{BH,\star}$, are almost devoid of (host) galaxies matching in mass, since they are determined only on small areas of the sky.

As a consequence the SMBH luminosity functions at $z=6$ only span 2\,dex of dynamic range, and part of this is Eddington ratio $\lambda$ and not \MBH. However, the assumption of a Schechter or double power law shape is motivated by a lower-$z$ success of describing data: studies of quasar LFs at lower redshifts paint a picture that strongly suggests a continuum of functional form of the LF, with only a slow evolution in each faint- and bright-end slopes (see e.g.\ comparisons in S23). So using this shape also to extrapolate to some lower $L$ seems valid, and the relative `flatness' of the LF at the faint end points to an actual lack of many observable UV-visible quasars at these luminosities, and not the current limits of observational facilities. With this interpretation the lower-$L$ cutoff is -- as discussed -- not strongly impacting the results due to the flat faint end, as are the details of how this relation evolves from higher redshifts. It simply seems as there are not very strong contributions by the lower-$L$ UV-luminous AGN and quasars, that we discuss here, to the overall SMBH mass buildup in the first billion years.

On the other hand, when judged by the stellar mass functions and \MBHMS\ scaling relations, the assumption of the scaling relations' validity might either break down at $10^7$, $10^6$, $10^5$\,\Msun, or might still hold towards lower masses. The former would imply that SMBHs below a certain mass have a formation path different from higher mass SMBHs -- without regulating feedback between galaxy and SMBH, or without having partaken in `cosmic averaging' \citep{jahnke2011}. If such low-mass SMBHs were on average less massive than expected from the \MBHMS-relation then they might not contribute much to $\rho_\mathrm{BH,\star}$ at $z=6$.

However, if the scaling relations correctly predict such a lower-\MBH\ population, this would imply that this substantial population of lower mass SMBHs would not grow strongly in a phase that we could observe as UV luminous AGN -- because they do not strongly show up in the (extrapolated) luminosity function. This population would have to grow differently. 

That said, we now want to evaluate the two sides of the Sołtan argument: How much of the SMBH mass at $z=6$ can we observe forming in UV-luminous quasars at $z>6$?
At this point we have picked the galaxy MF with the lowest resulting stellar mass  \citep{weibel2024}, the \MBHMS\ scaling relation with the lowest normalisation \citep{li2024a}, an up-to-date $z\sim6$ QLF \citep{schindler2023}, and conservative values for radiative efficiency and bolometric correction. The accreted SMBH mass density is $\rho_\mathrm{BH,acc} = 79\,\Msunm/\mathrm{Mpc}^3$. The resulting SMBH mass density at $z=6$ as inferred from the galaxy MF is $\MBHDsm(>10^6\,\Msunm) = 770$\,\Msun\,Mpc$^{-3}$ and $\MBHDsm(>10^5\,\Msunm) = 1530$\,\Msun/Mpc$^{-3}$.

Clearly, both choices of mass-cutoffs produce a $\MBHDsm$ that is at least an order of magnitude larger than the SMBH mass density we see build up from accretion in quasars. Is there any specific choice of mass-cutoff we should prefer for this comparison? This might be the wrong question, unless there is a reason, as said above, that the \MBHMS\ scaling relations should become invalid already above $10^5$ or $10^6$\,\Msun. It boils down to whether we would or would not have to add many quasars for consideration on the accretion side. But there, as shown, the extrapolation of the luminosity to lower and lower luminosities -- and hence SMBH growth -- does not increase the total mass substantially. Even though SMBHs with masses below $10^7$\,\Msun\ should clearly exists, under the assumption of the scaling relations holding towards lower masses as they do at later cosmic times, they would have to grow without much visible UV radiation. So no matter the mass integration cut-off choice, this comparison clearly states that the majority of SMBH mass buildup did not take place in these UV-luminous quasars from which we currently try to draw most of the detailed information about early SMBH formation.

\subsection{How much SMBH growth is obscured?}
\label{sec:soltanevaluation}

But how much of SMBH growth is really obscured? What is the state of knowledge so far? While this is the first time $\MBHDsm$ could be estimated, there exist previous estimates for $\rho_\mathrm{BH,acc}$. \citet{shen2020} construct a bolometric QLF from $z=0$ to 7, using X-ray, UV, visible, and infrared survey data, attempting to create a model that will fit all wavelengths at all redshifts. What is important, they include obscuration by gas and dust and include knowledge from mostly $z\le5$ X-ray surveys on fractions of absorbed and compton-thick quasars. They also provide an estimate of $\rho_\mathrm{BH,acc}$, as integrated since $z=10$ -- which should cover the majority of SMBH growth in the early Universe. Their figure 12 shows a value of $\rho_\mathrm{BH,acc}(>10^{43}\,\mathrm{erg/s})\sim 470\,\Msunm/\mathrm{Mpc}^3$ for $z=6$, a value substantially larger than our estimate. 

What their number and their bolometric QLF contain are factors accounting for obscuration and light attenuation both from hydrogen (for absorption in the X-ray regime) and derived dust extinction in the UV--visible wavelength regime. Their obscuration numbers for the early Universe are extrapolated from lower redshifts as data only reached to $z=5$, but they assume that of the set of Compton-thin quasars, 55--80\% see substantial absorption in the X-rays ($22<\log (N_\mathrm{H}/\mathrm{cm}^2)<24$ in hydrogen column surface density), and the numbers of Compton-thick quasars is $\sim$\,40\% relative to the Compton Thin population. 

Converted to dust extinction, \citet{shen2020} use $(A_\mathrm{B}/N_\mathrm{H}) = 8.47\times10^{-22}$\,cm$^{-2}$ and for a wavelength of 1450\,\AA\ -- where the QLF we are using is based on -- $A_\mathrm{1450\,\textnormal{\AA}}/A_\mathrm{B}$ ranges from $\sim$\,1.0 to $\sim$\,2.2, for dust models with $R(V)$ from 5.5 to 2.5 \citep{gordon2023}. This implies extinction for the `absorbed' class of quasars already exceeding 10\,mag which would clearly remove all absorbed and Compton-thick quasars from being observable at 1450\,\AA, and likely some fraction below $N_\mathrm{H}=10^{22}$\,cm$^{-2}$ as well. Overall, this will make a fraction of 70--90\% of all quasars entering the SMBH mass function of \citet{shen2020} unobservable at 1450\,\AA, explaining the factor five difference to our quasar-accreted mass estimates. 

What is interesting now is that \citet{shen2020} {\em still} predict SMBH mass densities substantially below the most conservative $\MBHDsm$ value we have estimated -- and their lower luminosity integration bound of $10^{43}\,\mathrm{erg/s}$ corresponds to SMBHs accreting at Eddington ($\lambda=1$) with $\MBHm \sim 10^5\,\Msunm$\footnote{At $\lambda=1$ these SMBHs would grow from $z=7.5$ to $z=6$ by two orders of magnitude in mass.}. So including SMBHs down to this mass on our side would increase the difference to the \citet{shen2020} estimates to a factor of 1530/470=3.3. We would still be missing many more growing SMBHs than they include in their integrated unobscured plus obscured accretion estimate. 
\medskip

What are the consequences? Our 1450\,\AA\ QLF-based $\rho_\mathrm{BH,acc} = 79\,\Msunm/\mathrm{Mpc}^3$ is a factor 10 smaller than the galaxy MF-based $\MBHDsm(>10^6\,\Msunm) = 770$\,\Msun/Mpc$^3$, which represents our most conservative set of assumptions. It could be justified to integrate further down, and with a slightly higher normalisation of the \MBHMS\ scaling relations this discrepancy would only grow stronger. It is clear: the vast majority of SMBH mass we infer for $z=6$ is not being accreted into SMBHs by the UV-luminous quasars that we observe at $z\ge6$, or the extrapolation of the observed QLF to fainter quasars. Even including the known obscured quasars seems likely missing a substantial part of the SMBH mass growth.

At the same time, this population of accreting SMBHs is at the core of our diagnostics of early SMBH growth: except for a few notable exceptions \citep[e.g.][]{suh2024} optically unobscured quasars are the only objects for which we can reasonably well estimate \MBH\ and specific accretion rates, that is the masses of a galaxy's SMBH and its growth rate. The whole observational anchoring of the `black hole seed' conundrum for the earliest Universe comes from the discrepancy of measured \MBH\ in individual $z>6$ quasars and their instantaneous specific accretion rates not systematically exceeding the Eddington growth rate \citep[e.g.][]{fan2023}. The `seed problem' only arises from the assumption that these SMBH masses at $z>7$ need to be reached with Eddington-limited growth as observed for this specific population of UV-selected broad-line quasars -- we will discuss this further below. While we already knew about a substantial population of SMBHs with dust-obscured growth, we do not know their masses and hence not their specific accretion rates.

Our calculations and the discrepancy between the two mass densities suggest something very different: viewed from a perspective of rest-frame UV--visible wavelength surveys the majority of SMBH growth appears in the `dark' and we can not extract detailed growth-diagnostic on it. This is qualitatively consistent with both the above higher inferred `bolometric' \MBHD-value by \citet{shen2020}, as well as predictions of the overall expectation of quasar extinction in the X-rays \citep[][]{vito2018}. Most quasars at $z>6$ will have either substantial amounts of gas and dust around their SMBHs, or throughout their galaxy, fuelling substantial star formation, or even both.

\subsection{Eddington limit and growth timescales}
\label{sec:timescales}

\begin{figure}[htb]
   \includegraphics[width=1.0\columnwidth]{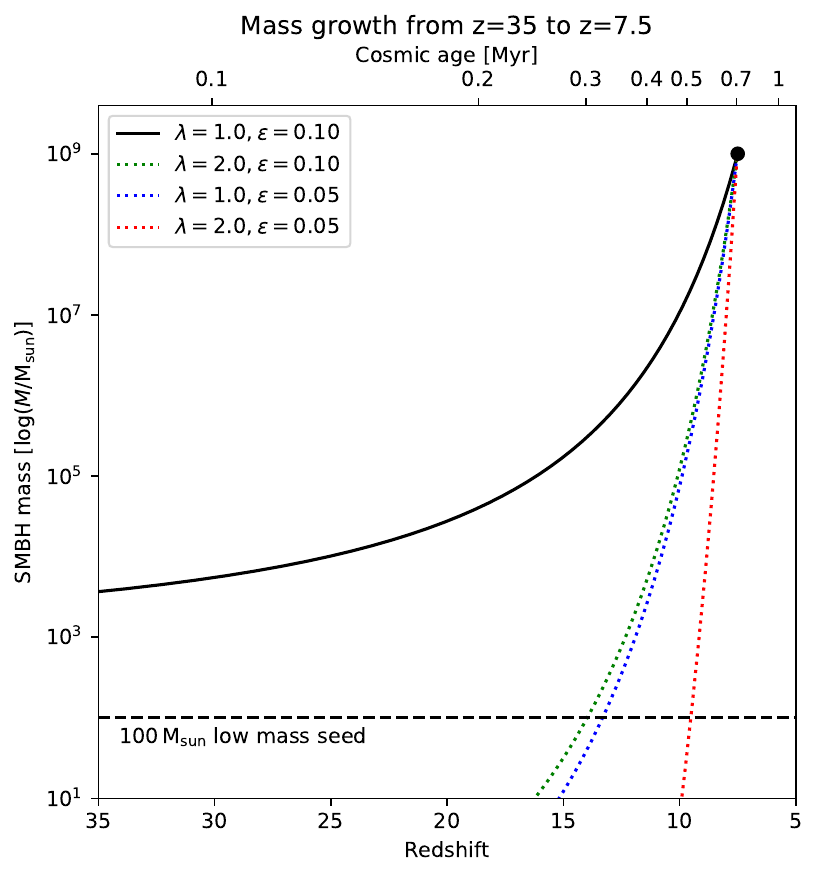}
    \caption{Growth of SMBH mass for a quasar with $\MBHm=10^9$\,\Msun\ at $z=7.5$. The solid black line shows growth with $\lambda=1.0$ and $\varepsilon=0.10$. The dotted lines show the growth if only the final $10^{6.5}$\,yr take place with these parameters -- see Sect.~\ref{sec:timescales}, this time interval is almost fully contained in the large black dot -- but with higher $\lambda$ or lower $\varepsilon$ or both earlier than that. While the final phase with $\lambda=1.0$ and $\varepsilon=0.10$ grows $\sim$\,7\% of the final mass, the growth before this time could obviously be much faster with different value combinations. It is solely our extrapolation from this observable final UV-luminous phase to earlier times that sets constraints preventing direct growth from low-mass SMBH seeds.
    \label{fig:massevo}}
\end{figure}

While dust already shields this SMBH growth, or at least its detailed diagnostics, from our view, as mentioned this also blinds us to two essential parameter for the majority of SMBH growth phases or populations: the specific accretion rate $\lambda=L_\mathrm{bol}/L_\mathrm{Eddington}$, and the the radiative efficiency $\varepsilon$. All $z=6$ quasars that we observe and can diagnose accrete at or close to $\lambda=1$ -- but not much above it: \citet{fan2023} quote a median $\lambda=0.79$ and a range from 0.08 to 2.7. With an $\varepsilon=0.1$ for a thin accretion disk and $\lambda=1.0$ as observed, the $e$-folding timescale for SMBH mass is the Salpeter timescale with these values,
\begin{equation}
    t_e = 4.5\times10^8 \varepsilon/((1-\varepsilon)\lambda)\,\mathrm{yr},
\end{equation}
that is $t_e=50$\,Myr. This growth rate would not leave enough time for the 10$^9$\,\Msun\ SMBHs at $z=7.5$ to grow from black hole seeds less massive than 10$^5$ or 10$^4$\,\Msun\ at, say, $z=30$ \citep[e.g.][]{banados2018,yang2020}. However, this issue immediately disappears if we remove the requirements for these two values as set by the -- as we here claim non-representative -- UV quasars. With either $\lambda=2.0$ or $\varepsilon=0.05$, that is twice or half their currently assumed values, respectively, the growth times from 100\,\Msun\ to $1.5\times10^9$\,\Msun\ shrinks to 400\,Myr, corresponding to a starting point $z_\mathrm{seed}\sim14$, as shown in \cref{fig:massevo}. For accretion with both $\lambda=2.0$ and $\varepsilon=0.05$ this time decreases by another factor two to 200\,Myr or $z_\mathrm{seed}\sim9.5$.

But is this reasonable given that we see all these luminous, high-mass quasars at $z>6$? There are completely independent arguments on the duration of the `luminous' quasar accretion phase, that is visible to us and centrally largely unobscured, from an analysis of UV-luminous duty cycle estimates $\ll$\,1 from clustering \citep{eilers2024}, as well as from quasar proximity zones \citep[e.g.][]{davies2018,eilers2021,satyavolu2023}. The latter are the physical regions of initially un- or only partially ionized hydrogen around quasars in the early Universe which get ionized by the quasar's UV radiation. The physical distance to which this happens is typically constrained, and this allows a diagnostic of `quasar lifetime' from the light-travel time of the ionizing photons and time-dependent spatial growth of the ionized region. Again, this lifetime is the duration of unobscured accretion, where ionizing photons are not absorbed by an intervening medium between the accretion disk and the surrounding environment. 

The best estimates of these durations are all consistent with $t_\mathrm{prox}\lesssim 10^{6.5}$\,yr \citep[e.g.][]{davies2018,andika2020,duro2024} at $z\ge$6. This is a short period of time, given the total growth history. Since these quasars with very mature SMBHs of masses $10^8$ or $10^9$\,\Msun, it is clear that these SMBHs are close to the end of their growth history, given that we do not know of much more massive systems in the local Universe. This suggests that these proximity zone `quasar lifetimes' might very well describe a final, if not the final unobscured accretion phase. The growth of these very massive SMBHs must be in the process of tapering down, and the observed specific accretion rates do not need to be in any way representative for their earlier phases of growth.

If we then take this duration of unobscured accretion $t_\mathrm{prox}\sim 10^{6.5}$\,yr and compare it to the e-folding Salpeter timescale $t_e=50$\,Myr for $\varepsilon=0.1$ and $\lambda=1.0$, these SMBHs grew by $\sim$\,7\% in this luminous phase. This is directly consistent with the $\le$1:10 ratio of $\rho_\mathrm{BH,acc}$ to $\MBHDsm$ that we independently derived above based on those very same quasars.

All this provides strong evidence to a scenario in which SMBHs can start out from any light SMBH seed mass, grow in the extreme gas densities of the tumultuous early Universe at $z>10$ in a way that is not limited by the Eddington accretion rate, possibly with high accretion rate slim disks with substantially lower $\varepsilon<0.1$ \citep{abramowicz2013}. There is independent evidence for this also from quasar proximity zones \citep{davies2019}, and from at least one example at $z=4$ \citep{suh2024}. In this case the need for massive SMBH seeds would immediately disappear or rather morph into a question of sustained super-Eddington gas supply but would -- empirically from the $z\sim6$ point of view -- not be bound by $\lambda\le1$. At some point in time, in systems where the gas supply dwindles or the accretion geometry changes so that the accretion rate drops, part of the remaining gas might be blown away from the vicinity of the SMBH, the vicinity of the quasar both becomes unabsorbing for ionizing UV photons, and transparent for at least some lines of sight to us as observers, while at the same time potentially further impacting its future accretion reservoir by this feedback. All luminous and observable quasar phases would therefore be young and late phases, while $>90$\% of the SMBH growth had taken place before in obscured phases. These obscured growth phases could either be episodic \citep{li2023} or continuous. Since they would mostly take place in phases where the $\MBHm<10^8$\,\Msun, their radiative UV--visible wavelength emission that is reprocessed into the infrared would not appear anomalously high in surveys, even for high $\lambda$, since the luminosity scales with both $\lambda$ and \MBH\ -- and we can not diagnose either when solely knowing a quasar's infrared emission. 

What this would indeed predict though is an unexpectedly {\em numerous} population of infrared-luminous sources at high $z$, intrinsically powered by accreting SMBHs. This could be an explanation for the currently debated nature of the abundant early-Universe `Little Red Dots' that are being seen with JWST \citep[e.g.][]{greene2024,li2024b,akins2024,ananna2024,matthee2024,suh2024} and where the debate is ongoing whether these all contain AGN or not. Our analysis would lend weight to a substantial AGN population among them. If a fraction of their emitted light would stem from SMBH accretion processes, then this would also be sufficient to explain the $z=6$ SMBH mass volume density \citep{inayoshi2024}.


\section{Conclusions and summary}
\label{sec:conclusions}

We want to condense our finding into the following conclusions:

\begin{enumerate}
    \item It has now become possible to estimate the SMBH volume density $\MBHDsm$ at $z=6$ using stellar mass function and the first estimates of \MBHMS\ scaling relations; it is therefore possible to recreate the Sołtan argument at $z=6$.

    \item The UV--visible wavelength population of luminous quasars at $z\ge6$, that is the only one to provide estimates of \MBH\ and specific accretion rates in the early Universe, is responsible for $\le10$\% of $\MBHDsm$ at $z=6$.

    \item The remaining $\ge$90\% of SMBH volume density buildup must take place in quasars that are highly obscured in the UV--visible wavelength regime, therefore usually not allowing \MBH\ and Eddington ratio estimates. Prior estimates of total SMBH buildup that take into account the known obscured population might still miss a sizeable fraction of SMBH growth.

    \item The phase of $z\ge6$ quasar activity in luminous UV--visible light is likely very short-lived, only representing a late or final stage of SMBH growth. Independent evidence for this comes from diagnostics of quasar proximity zones, which limits luminous quasar phase lifetimes and radiative efficiencies.
    
    \item The `SMBH seed problem' to require massive SMBH seeds in the very early Universe is solely due to the assumption that the known UV-bright quasars represent the typical mode of SMBH growth in the early Universe, with standard thin accretion disks of radiative efficiency $\varepsilon\sim0.1$ and specific accretion rate $L/L_\mathrm{Eddington} = \lambda\sim1$. According to our analysis this phase is by far not representative of early SMBH growth.

    \item With $\ge$90\% of SMBH growth not taking place in these systems, the growth modes for the majority of \MBH\ likely differ from the ones currently observed. The seed problem disappears as an observational conundrum, since other, faster modes of growth in the gas-abundant earliest Universe can now be called on, well possible with substantially higher $\lambda$ and/or lower $\varepsilon$, potentially with slim disk accretion disks. 

    \item This implies that there need to be large populations of highly obscured quasars at $z\ge6$, even larger than known at later cosmic times. These will not be accessible through UV--visible selection due to fully obscured quasar nuclei, but likely need to be searched for in the near- or mid-infrared. Their volume number density should be substantially higher than that of the unobscured quasar population. These could well be the `Little Red Dots' seen and currently being analysed with JWST and X-ray data.

    \item This limits the impact of quasar searches with new surveys at $z>7$ \citep[e.g.\ {\em Euclid}, Roman, and Rubin,][]{barnett2019,tee2023}, as the quasars found will mostly represent late or final stages of SMBH growth. They will be very useful for quasar absorption line studies and diagnostics of extreme \MBH--\MS\ values. However they will have a limited impact to diagnose overall SMBH growth in the early Universe, mainly to constrain the final stages of SMBH buildup in the $z=6$ quasar population.
\end{enumerate}

\section*{Acknowledgments}
We thank would like to thank John Silverman, Coryn Bailer-Jones, James Davies, Frederick Davies, Anna-Christina Eilers, Eduardo Bañados, Hans-Walter Rix, and Annalisa Pillepich for very helpful input, criticism, checks, and discussions.

This work made use of Astropy\footnote{https://www.astropy.org}, a community-developed core Python package and an ecosystem of tools and resources for astronomy \citep{astropy2022}.

\bibliography{darkaccretion}
\bibliographystyle{mnras}

\end{document}